\documentclass[a4paper,11pt]{article}

\usepackage{amsmath}
\usepackage{amssymb}
\usepackage{color}
\usepackage{cite}
\usepackage{hyperref}

\makeatletter
\@addtoreset{equation}{section}
\renewcommand{\theequation}{\thesection.\@arabic\c@equation}
\makeatother
\makeatletter
\renewcommand\appendix{\par
  \setcounter{section}{0}%
  \setcounter{subsection}{0}%
  \gdef\thesection{Appendix \@Alph\c@section }
  \renewcommand{\theequation}
  {\Alph{section}.\arabic{equation}}
}
\makeatother

\def \be {\begin{equation}}
\def \ee {\end{equation}}
\def \ba {\begin{array}}
\def \ea {\end{array}}
\def \bea{\begin{eqnarray}}
\def \eea{\end{eqnarray}}

\def \a {\alpha}
\def \b {\beta}
\def \g {\gamma}
\def \G {\Gamma}

\def \D {\Delta}
\def \e {\epsilon}

\def \m {\mu}
\def \n {\nu}

\def \l {\lambda}
\def \L {\Lambda}
\def \s {\sigma}

\def \o {\omega}
\def \O {\Omega}

\def \p {\partial}
\def \f {\frac}

\def \nn {\nonumber}

\def \ma {\mathcal}

\def \lt {\left}
\def \rt {\right}

\def \ra {\rightarrow}

\def \sr {\sqrt}
\def \td {\tilde}
\def \hs {\hspace}

\def \inf {\infty}
\def \dd {\mathrm{d}}
\def \tph {{\tilde\phi}}
\def \tps {{\tilde\psi}}
\def \hph {{\hat\phi}}
\def \hps {{\hat\psi}}
\def \ho {{\hat\omega}}

\def \hatt {{\hat t}}
\def \hatr {{\hat r}}

\title{\textbf{Holographic Descriptions of Black Rings}}
\author{
Bin Chen$^{1,2}$\footnote{bchen01@pku.edu.cn}\,
and
Jia-ju Zhang$^{1}$\footnote{jjzhang@pku.edu.cn}
}
\date{}

\begin{document}

\maketitle

\begin{center}
{\it
$^{1}$Department of Physics and State Key Laboratory of Nuclear Physics and Technology, Peking University, Beijing 100871, P.R. China\\
\vspace{2mm}
$^{3}$Center for High Energy Physics, Peking University, Beijing 100871, P.R. China\\
}
\vspace{10mm}
\end{center}

\begin{abstract}

In this paper, we investigate the holographic descriptions of two kinds of black rings, the neutral doubly rotating black ring and the dipole charged black ring. For generic nonextremal black rings, the information of holographic CFT duals, including the central charges and left- and right-moving temperatures, could be read from the thermodynamics at the outer and inner horizons, as suggested in \cite{Chen:2012mh}. To confirm these pictures, we study the extreme black rings in the well-established formalism. We compute the central charges of dual CFTs by doing asymptotic symmetry group analysis in the stretched horizon formalism, and find exact agreements. Moreover, we study the superradiant scattering of a scalar field off the near-extremal black rings and obtain the scattering amplitudes, which are in good match with the CFT predictions.

\end{abstract}

\baselineskip 18pt
\thispagestyle{empty}

\newpage

\section{Introduction}

The Kerr/CFT correspondence states that a four-dimensional Kerr black hole could be holographically described by a two-dimensional conformal field theory (CFT) with equal central charges in left- and right-moving sectors. It is well-understood in the extreme case, which has only one horizon and vanishing Hawking temperature. In \cite{Guica:2008mu}, the asymptotic symmetry group (ASG) of the near horizon geometry of extreme Kerr black hole (NHEK) was analyzed in the Barnich-Brandt-Compere (BBC) formalism \cite{BBC}. Under suitable falloff conditions, the ASG of NHEK geometry form  a Witt algebra (Virasoro algebra without central charge) classically, whose quantum version gives the Virasoro algebra with a central charge $c_L=12J$ where $J$ is the angular momentum of Kerr black hole. Moreover, it was found later that there exists another copy of Virasoro algebra with the same central charge \cite{Matsuo:2009sj, Castro:2009jf}, which gives the right-mover of the CFT. In the extreme case, the excitations in right-moving sector of dual CFT are suppressed such that the right-moving temperature is vanishing, in accordance with the fact that the Hawking temperature of extreme Kerr is vanishing. The similar ASG analysis has been generalized to extreme Kerr and Kerr-Newman black holes in various dimensions\cite{Lu:2008jk,Hartman:2008pb} to find their holographic duals. Actually such kind of analysis can give only the central charge in the left-mover of dual CFT. However, it is believed that the central charge in the right-mover should take the same value. Further support of the Kerr/CFT correspondence for extreme Kerr black hole are found in the studies of the superradiant scattering processes\cite{Bredberg:2009pv,Cvetic:2009jn,Hartman:2009nz,Chen:2010ni}. For the nonextreme black holes, the Kerr/CFT correspondence is mainly based on the existence of the hidden conformal symmetry in the low-frequency scattering in the near horizon region\cite{Castro:2010fd}. From the periodic identification on the conformal coordinates, the dual temperatures are read out, with which the macroscopic Bekenstein-Hawking entropy is reproduced and  the scattering amplitudes are found to be in consistent with the CFT predictions\cite{Chen:2010xu}. However there is no derivation on the central charges of dual CFT. It is believed that the central charges should take the similar form as those of extreme case, provided that the central charges could be written in terms of ``quantized" quantities like angular momenta and $U(1)$ charges. For further details on Kerr/CFT correspondence and more complete references, see the nice reviews  \cite{KerrCFT}.

Very recently, it was found that there is close relation between the Kerr/CFT correspondence and black hole thermodynamics. It has been found for a long time that the existence of holographic descriptions of black holes is related to the fact that the area products of the outer and inner horizons are independent of these black hole masses \cite{Larsen:1997ge,Cvetic:1997uw,Cvetic:1997xv,Cvetic:2010mn}.
In \cite{Hartman:2008pb}, it was shown that the CFT left-moving temperature could be got from the thermodynamics of the black hole in the extremal case.
Moreover, in a remarkable paper \cite{Cvetic:2009jn}, from the thermodynamics of outer and inner horizons of multi-charged Kerr black hole, the thermodynamics in the left- and right-moving sectors of the dual CFT was discussed, from which the central charges were reproduced in the extreme limit. In our recent work \cite{Chen:2012mh}, we investigated more carefully the thermodynamics of the inner horizon and its implication on the holographic description of the black hole. Under reasonable assumption, we proved that the first law of thermodynamics of the outer horizon always indicates that of the inner horizon. As a result, the fact that the area product being mass-independent is equivalent to the relation $T_+S_+=T_-S_-$, with $T_\pm$ and $S_\pm$ being the Hawking temperatures and the entropies of the outer and inner horizon respectively. We showed that once the relation $T_+S_+=T_-S_-$ is satisfied, the central charges of two sectors in the dual CFT must be same. Furthermore from the thermodynamics relations, we read directly the dimensionless temperatures of microscopic CFT, which are in exact agreement with the ones obtained from hidden conformal symmetry in the low frequency scattering off the black holes, and then determine the central charges. This method works well in well-known cases in Kerr/CFT correspondence, and reproduce successfully the holographic pictures for 4D Kerr-Newman and 5D Kerr black holes. It seems that all the information on the holographic CFT dual for the black holes is encoded in the black hole thermodynamics. It is interesting to see if this is true for other kinds of black objects.

Black ring is a special kind  of asymptotically flat five-dimensional black object. Different from the usual five-dimensional Myers-Perry black hole, whose horizon has the topology of $S^3$, the black ring horizon has the topology of $S^1 \times S^2$.\footnote{In this paper we denote the angle of $S^1$ as $\psi$ and an angle of $S^2$ as $\phi$.} The first black ring was discovered in \cite{Emparan:2001wn}, which is a neutral singly rotating black ring along $S^1$. Subsequently, various kinds of electric and dipole charged black rings \cite{Elvang:2003yy,Emparan:2004wy,Elvang:2004rt,Elvang:2004ds,Elvang:2004xi} and neutral doubly rotating black ring \cite{Pomeransky:2006bd} were constructed. A general charged black ring could have nine parameters, i.e.\! the mass $M$, two angular momenta $J_\psi,J_\phi$, three electric charges $Q_{1,2,3}$ and three dipole charges $q_{1,2,3}$. However, no such solution has been constructed explicitly. The black rings are similar to the black hole in many aspects: the area law of the entropy, thermodynamics, et.al. But the black rings present the first example of non-uniqueness of black objects in higher dimensional Einstein gravity. More details on black rings can be found in the reviews \cite{Emparan:2006mm,Emparan:2008eg} and references therein.

Just like that for multi-charged BPS black holes, there have been microstates countings for various extremal black rings by embedding the black rings into  M5-M2 branes system in M-theory. These include the dipole black ring \cite{Emparan:2004wy}, BPS black ring \cite{Bena:2004tk,Cyrier:2004hj}, neutral doubly rotating black ring \cite{Reall:2007jv}, a kind of electric and dipole charged black ring \cite{Emparan:2008qn} and even a general charged  nearly extremal black ring without explicit construction\cite{Larsen:2005qr}.
After the appearance of Kerr/CFT, there were trials of reproducing the entropy of black ring using ASG analysis. For examples, the BPS black ring in 5D $N=2$ supergravity has been studied from different point of views \cite{Loran:2008mm,Goldstein:2011jh}, including Brown-Henneaux ASG approach\cite{Brown:1986nw}, c-extremization approach \cite{Kraus:2005vz} and the entropy function formalism \cite{Sen:2005wa}.
One important feature of this supersymmetric black ring is that their near horizon geometry is locally AdS$_3$, which is essential to make use of  c-extremization approach. In this case, the near horizon geometry is very different from the NHEK geometry and also different from the those of black rings studied in this paper.

In this paper, we would like to study the holographic descriptions of black rings in the framework of Kerr/CFT correspondence and inner horizon thermodynamics. Unlike the Kerr black hole, the CFT duals for general nonextremal black rings are unknown. It seems that we can get the dual CFT temperatures of a nonextremal black ring from the hidden conformal symmetry. But it appears that the scalar equation under the general black ring background cannot be variable separated, so the hidden conformal symmetry in the radial equation is missing. Nevertheless, it is possible to apply the thermodynamics method to set up the correspondence. In fact, in \cite{Chen:2012mh} we applied the thermodynamic method to the neutral doubly rotating black ring found in \cite{Pomeransky:2006bd}, and found that  there is only one CFT dual picture with central charge $c_L^\phi=c_R^\phi=12J_\phi$. In this paper we will verify this result using the conventional ASG methods. Also we will use the method to analyze the the dipole black ring found in \cite{Emparan:2004wy}, and corroborate the effectiveness of the thermodynamics method.

The remaining part of this paper is as follows. In Section~\ref{DRBR}, we review the CFT dual of the doubly rotating black ring and verify the dual by analyzing the near horizon geometry of extreme black ring. In Section~\ref{DBR}, we study the dipole black ring and set up its CFT dual. In Section~\ref{SC}, we discuss the superradiant scalar scattering off the extreme black rings and find that the scatter amplitudes are in agreement with CFT prediction. We end with conclusion and discussion in Section \ref{CONC}.

\section{Neutral Doubly Rotating Black Ring}\label{DRBR}

In this section, we firstly review the properties of the neutral doubly rotating black ring, and then show the CFT dual for the nonextremal black ring  from the horizons' thermodynamics. Finally we verify the black ring/CFT duality using the conventional ASG analysis.

\subsection{Metric}

The  neutral doubly rotating black ring was discovered in \cite{Pomeransky:2006bd}. It has the metric of the form \footnote{Here we use the convention that is a little different from that in \cite{Pomeransky:2006bd}, we use a mostly plus signature, and our coordinates $\phi$, $\psi$ are respectively $\psi$, $\phi$ therein.},
\bea
ds^2&=&-\frac{H(y,x)}{H(x,y)}(dt+\O)^2+\frac{F(y,x)}{H(y,x)}d\psi^2-2\frac{J(x,y)}{H(y,x)}d\psi d\phi\nn\\
&&-\frac{F(x,y)}{H(y,x)}d\phi^2+\frac{2R^2H(x,y)}{(x-y)^2(1-\n)^2}\left(\frac{dx^2}{G(x)}-\frac{dy^2}{G(y)}\right),\label{doubler}
\eea
where the coordinates ranges are $-\inf<t<\inf$, $-1<x<1$, $-\inf<y<-1$, and the angles $\psi,\phi$ have periods of $2\pi$. In (\ref{doubler}), there are functions
\bea
\O&=&-\f{2R\l\sr{(1+\n)^2-\l^2}}{H(y,x)} \lt[  -\f{1+y}{1-\l+\n}\lt( 1+\l-\n+x^2 y \n(1-\l-\n)+2\n x(1-y) \rt)d\psi \rt.   \nn\\
&&   \lt.  \phantom{-\f{2R\l\sr{(1+\n)^2-\l^2}}{H(y,x)} \lt[\rt.}
                                               +(1-x^2)y\sr{\n}d\phi    \rt],   \nn\\
G(x)&=&(1-x^2)(1+\l x+\n x^2),  \nn\\
H(x,y)&=&1+\l^2-\n^2+2\l\n(1-x^2)y+2x\l(1-y^2\n^2)+x^2y^2\n(1-\l^2-\n^2),  \nn\\
J(x,y)&=& \f{2R^2(1-x^2)(1-y^2)\l\sr{\n}}{(x-y)(1-\n)^2} \lt[ 1+\l^2-\n^2+2(x+y)\l\n-xy\n(1-\l^2-\n^2) \rt],  \nn\\
F(x,y)&=&\f{2R^2}{(x-y)^2(1-\n)^2} \lt\{ G(x)(1-y^2) \lt[ \lt( (1-\n)^2 -\l^2\rt)(1+\n)+y\l(1-\l^2+2\n-3\n^2) \rt] \rt.  \nn\\
&&    \lt. + G(y) \lt[ 2\l^2 +x\l \lt( (1-\n)^2+\l^2 \rt) +x^2  \lt( (1-\n)^2 -\l^2\rt)(1+\n) +x^3 \l(1-\l^2-3\n^2+2\n^3)  \rt.\rt. \nn\\
&&    \lt. \lt.  \phantom{+ G(y)} -x^4(1-\n)\n (-1+\l^2+\n^2) \rt]    \rt\}.
\eea
The black ring is characterized by three parameters $R,\n,\l$, with $R$ having dimension of length and $\n,\l$ being dimensionless.  There are constrains $0\leq\n<1$ and $2\sr{\n}\leq \l <1+\n$ in order to make the black ring regular. Following \cite{Chen:2012mh} we introduce the coordinate $r=-1/y$ with $0<r<1$, in terms of which the outer and inner horizons at $r_\pm$ are determined by $r^2-\l r+\n=0$. Both of the black ring horizons have the topology of $S^1 \times S^2$. The mass, two angular momenta, the entropy, and the Hawking temperature of the outer horizon are respectively
\bea \label{e15}
&& M=\f{3\pi R^2(r_+ + r_-)}{(1-r_+)(1-r_-)},  \nn\\
&& J_\psi=\f{2\pi R^3 (r_++r_-)(1+r_++r_--6r_+r_-+r_+^2r_-+r_+r_-^2+r_+^2r_-^2)}{(1-r_+)(1-r_-)(1-r_+r_-)^2}
          \sr{\f{(1+r_+)(1+r_-)}{(1-r_+)(1-r_-)}},  \nn\\
&& J_\phi=\f{4\pi R^3 (r_++r_-)}{(1-r_+r_-)^2}\sr{\f{r_+r_-(1+r_+)(1+r_-)}{(1-r_+)(1-r_-)}}, \nn\\
&& S_+=\f{8\pi^2 R^3 r_+(1+r_-)(r_++r_-)}{(1-r_+)(1-r_+r_-)^2},\hs{3ex} T_+=\f{(1-r_+r_-)(1-r_+)(r_+-r_-)}{8\pi R r_+(1+r_-)(r_++r_-)}.
\eea
When $\l^2=4\n$,  $r_+=r_-$,  the black ring becomes extremal, and then
\bea \label{e16}
&& M=\f{6\pi R^2 r_+}{(1-r_+)^2},  \nn\\
&& J_\phi=\f{8\pi R^3r_+^2}{(1+r_+)(1-r_+)^3},  \nn\\
&& J_\psi=\f{4\pi R^3 r_+(1+4r_++r_+^2)}{(1+r_+)(1-r_+)^3},  \nn\\
&& S=2\pi J_\phi, \hs{3ex}T=0.
\eea

\subsection{CFT from thermodynamics}

Using the thermodynamics of the nonextremal black ring, it was conjectured in \cite{Chen:2012mh} that there is a CFT dual for the doubly rotating black ring. The reader could see \cite{Chen:2012mh} for detailed analysis, or see the next section for the similar discussion on the dipole black ring. For a general nonextremal doubly rotating black ring, there is a dual $J_\phi$ picture in which the CFT  has respectively the left-moving,  right-moving temperatures and the central charges
\bea
&&T_L^\phi=\f{(1-r_+r_-)(r_++r_-)}{4\pi\sr{r_+r_-(1-r_+^2)(1-r_-^2)}},  \nn\\
&&T_R^\phi=\f{(1-r_+r_-)(r_+-r_-)}{4\pi\sr{r_+r_-(1-r_+^2)(1-r_-^2)}}.  \nn\\
&&c_L^\phi=c_R^\phi=12J_\phi.
\eea
The entropy of the nonextremal black ring (\ref{e15}) could be reproduced by using the Cardy formula
\be
S=\f{\pi^2}{3}(c_L^\phi T_L^\phi+c_R^\phi T_R^\phi).
\ee
Taking the extremal limit, we get the CFT dual for the extremal black ring with the temperature and the central charge
\be \label{e17}
T_L^\phi=\f{1}{2\pi}, ~~~ c_L^\phi=12J_\phi,
\ee
Similar to the Kerr black hole the excitations in the right-moving sector is suppressed in the extreme limit such that $T_R^\phi=0$.

Note that there is no $J_\psi$ picture corresponding to the rotation along $\psi$. This  is in accord with the fact that for the original black ring with only the rotation along $\psi$ there is no regular extremal limit and therefore no CFT dual corresponding to such $U(1)$ symmetry.

\subsection{CFT from ASG analysis}

In this section we re-derive the extremal black ring/CFT correspondence (\ref{e17}) using conventional ASG analysis. There are two equivalent formalisms, the BBC formalism \cite{Guica:2008mu,BBC} and the stretched horizon formalism \cite{C1,C2}, to do such analysis. Here we just follow the treatment in \cite{Chen:2011wm}.

In coordinates $(t,r,x,\phi,\psi)$, the metric of the black ring could be written in the form
\be
ds^2=-N^2 dt^2+g_{rr}dr^2+g_{xx}dx^2+g_{mn}(d\phi^m+N^m dt)(d\phi^n+N^n dt),
\ee
with $m,n=(\phi,\psi)$, and
\bea
&& N^\phi=\f{g_{t\phi}g_{\psi\psi}-g_{t\psi}g_{\phi\psi}}{g_{\phi\phi}g_{\psi\psi}-g_{\phi\psi}^2},  \nn\\
&& N^\psi=\f{g_{\phi\phi}g_{t\psi}-g_{\phi\psi}g_{t\phi}}{g_{\phi\phi}g_{\psi\psi}-g_{\phi\psi}^2},  \nn\\
&& N^2=-g_{tt}+g_{mn}N^m N^n.
\eea
Following \cite{Chen:2011wm} and for an extremal black ring, we expand all the quantities near the horizon and define
\bea
&& N^2=(r-r_+)^2 f_1^2+\ma O(r-r_+)^3,  \nn\\
&& g_{rr}=\f{f_2^2}{(r-r_+)^2} +\ma O(r-r_+)^{-1},  \nn\\
&& N^m=-\O_+^m+(r-r_+)f_3^m+\ma O(r-r_+)^2,  \nn\\
&& f_4=f_2/f_1, ~~~ f^m=f_3^m f_4,  ~~~ m=\phi,\psi .
\eea
Explicit calculation shows that
\bea
&& f_1^2=\f{(1-r_+)^2[(1+r_+^2)(1+x^2)+4r_+ x]}{8r_+^2(1+r_+)^2(1+r_+x)^2},  \nn\\
&& f_2^2=\f{2R^2 r_+^2[(1+r_+^2)(1+x^2)+4r_+ x]}{(1-r_+^2)^2(1+r_+x)^2},  \nn\\
&& f_3^\phi=\f{(1-r_+)^2}{4R r_+^2}, ~~~ f_3^\psi=0,  \nn\\
&& f_4=\f{4Rr_+^2}{(1-r_+)^2}, ~~~ f^\phi=1, ~~~ f^\psi=0.
\eea
We define the new coordinates
\bea
&& t \ra \f{f_4}{\e }t,  \nn\\
&& r \ra r_++\e r,  \nn\\
&& \phi \ra  \phi+\f{\O_+^\phi f_4}{\e}t, \nn\\
&& \psi \ra  \psi+\f{\O_+^\psi f_4}{\e}t, \nn
\eea
and take the limit $\e \ra 0$, then we could get the near horizon geometry of the extremal black ring
\be \label{e1}
ds^2=\a(x) \lt( -r^2 dt^2+\f{dr^2}{r^2} \rt)+\b(x)dx^2 +\g_{mn}(x)(d\phi^m+f^m r dt)(d\phi^n+f^n r dt),
\ee
where
\bea
&& \a(x)=f_2^2=\f{2R^2 r_+^2[(1+r_+^2)(1+x^2)+4r_+ x]}{(1-r_+^2)^2(1+r_+x)^2},  \nn\\
&& \b(x)=g_{xx}|_{r=r_+}=\f{2R^2 r_+^2[(1+r_+^2)(1+x^2)+4r_+ x]}{(1-r_+^2)(1+r_+x)^4(1-x^2)},  \nn\\
&& \g_{\phi\phi}=g_{\phi\phi}|_{r=r_+}=\f{8R^2r_+^2(1-x^2)}{(1-r_+^2)[(1+r_+^2)(1+x^2)+4r_+ x]}, \\
&& \g_{\phi\psi}=g_{\phi\psi}|_{r=r_+}=\f{4R^2r_+(1+4r_++r_+^2)(1-x^2)}{(1-r_+^2)[(1+r_+^2)(1+x^2)+4r_+ x]}, \nn\\
&& \g_{\psi\psi}=g_{\psi\psi}|_{r=r_+}= \f{2R^2h(r_+,x)}{(1+r_+)(1-r_+)^2[(1+r_+^2)(1+x^2)+4r_+ x]},    \nn
\eea
with the function
\bea \label{h}
&& h(r_+,x)=3+13r_++18r_+^2-2r_+^2-r_+^4+r_+^5+8r_+(1+r_+)^3x   \nn\\
&& \phantom{g(r_+,x)=} +(1-r_+-2r_+^2+18r_+^3+13r_+^4+3r_+^5)x^2.
\eea
The area of the horizon at $r=0$ is
\be
A_+=4\pi^2 \int_{-1}^1 dx \sr{\b(x)\g(x)}=8\pi J_\phi,
\ee
where
\be
\g(x)= \g_{\phi\phi}\g_{\psi\psi}-\g_{\phi\psi}^2=\f{32R^4r_+^2(1+r_+)(1-x^2)}{(1-r_+)^3[(1+r_+^2)(1+x^2)+4r_+ x]}.
\ee

The geometry (\ref{e1}) is a fibred-AdS$_2$ space for fixed $x$, but the fibre is not simply a $S^1$. Such geometry appears also in the near-horizon geometry of extremal 5D Kerr or uplifted Kerr-Newman black holes.
It was demonstrated in \cite{Chen:2011wm} that for the geometry (\ref{e1}), there is always a CFT dual from ASG analysis, no matter in BBC formalism \cite{Guica:2008mu,BBC}, or in the stretched horizon formalism \cite{C1,C2}. In the doubly rotating black ring case, the CFT has the central charge
\be
c_L^\phi=\f{3f^\phi A_+}{2\pi}=12J_\phi,
\ee
and from Frolov-Thorne vacuum the temperature of the CFT is
\be
T_L^\phi=\f{1}{2\pi f^\phi}=\f{1}{2\pi}.
\ee
Because that $f^\psi=0$, there is no $J_\psi$ picture.

It is remarkable in \cite{Reall:2007jv}, a microstates counting of a doubly rotating extremal black ring has been proposed by mapping the black ring to a black string. Obviously, the treatment there is very different from ours.

\section{Dipole Black Ring}\label{DBR}

In this section we apply the analysis in the last section to the dipole black ring. We find that the $J_\psi$ picture appears due to the existence of the dipole charge.

\subsection{Metric}

There are general solutions with three unequal dipole charges, $(q_1,q_2,q_3)$, but for simplicity we just consider the case of equal dipole charges $q_1=q_2=q_3=q$. The dipole charged black ring was constructed in \cite{Emparan:2004wy}, and its metric is of the form
\bea
&& ds^2=-\f{F(y)H(x)}{F(x)H(y)} \lt( dt+C(\n,\l)R\f{1+y}{F(y)}d\psi \rt)^2  +\f{R^2}{(x-y)^2}F(x)H(x)H(y)^2 \nn\\
&& \phantom{ds^2=}
     \times \lt[ -\f{G(y)}{F(y)H(y)^3}d\psi^2-\f{dy^2}{G(y)}+\f{dx^2}{G(x)}+\f{G(x)}{F(x)H(x)^3}d\phi^2 \rt], \label{dipoler}
\eea
with the functions
\bea
&& F(\xi)=1+\l \xi, ~~~ G(\xi)=(1-\xi^2)(1+\n \xi),  \nn\\
&& H(\xi)=1-\m \xi, ~~~  C(\n,\l)=\sr{\l(\l-\n)\f{1+\l}{1-\l}}.
\eea
The coordinates $-1 \leq x \leq 1$, $-\infty < y\leq -1$.

There seems to be four parameters $(R,\l,\n,\m)$ in (\ref{dipoler}) to characterize the solution, but there is actually a redundancy: the relation
\be
\f{1-\l}{1+\l}=\lt( \f{1-\n}{1+\n} \rt)^2 \lt( \f{1-\m}{1+\m} \rt)^3
\ee
must be satisfied to have a regular spacetime. The parameter $R$ has the dimension of length, and the other dimensionless parameters are in the ranges
\be
0<\n \leq \l<1, ~~~ 0 \leq \m < 1.
\ee
The black ring has outer and inner horizons located at $y_+=-1/\n$ and $y_-=-\infty$ respectively. The angles $(\phi,\psi)$ have the periods
\be
\D\phi=\D\psi=2\pi \f{(1+\m)\sr{(1+\m)(1-\l)}}{1-\n}.
\ee

The dipole black ring (\ref{dipoler}) is characterized by three physical quantities,  the mass $M$, the angular momentum $J_\psi$ along $S^1$ , and the dipole charge $q$. For convenience we scale $q \to \f{(2\pi)^{1/3}}{\sr{3}}q$, and so $\Phi_\pm \to \f{\sr{3}}{(2\pi)^{1/3}} \Phi_\pm$. The re-scaled $q$ is the number of $M5$ branes from M-theory point of view, and thus must be an integer. The mass, the angular momentum and the dipole charge of the ring are respectively
\bea
&& M=\f{3\pi R^2}{4}\f{(1+\m)^2(\l+\m)}{1-\n},  \nn\\
&& J_\psi=\f{\pi R^3}{2} \f{(1+\m)^4}{(1-\n)^2} \sr{(1+\m)\l(1+\l)(\l-\n)},  \nn \\
&& q=(2\pi)^{1/3} R \f{1+\m}{1-\n} \sr{\f{\m(\m+\n)(1-\l)}{1-\m}},
\eea
and the Hawking temperature, the entropy, the angular velocity and dipole potential at the outer horizon are
\bea
&& T_+=\f{1}{4\pi R} \f{\n(1+\n)}{\m+\n} \sr{\f{1-\l}{(\m+\n)\l(1+\l)}} ,  \nn\\
&& S_+=2 \pi^2 R^3 \f{(1+\m)^3(\m+\n)}{(1+\n)(1-\n)^2}\sr{(\m+\n)\l(1-\l^2)},  \nn\\
&& \O_+^\psi=\f{1}{R} \f{1}{1+\m} \sr{\f{\l-\n}{(1+\m)\l(1+\l)}},  \nn\\
&& \Phi_+=\f{3\pi R}{2(2\pi)^{1/3}}(1+\m) \sr{\f{\m(1-\m)(1-\l)}{\m+\n}}.
\eea
The corresponding quantities at the inner horizon are \cite{Castro:2012av}
\bea
&& T_-=\f{1}{4\pi R} \f{\n}{\m} \sr{\f{1-\l}{\m(1+\l)(\l-\n)}},  \nn\\
&& S_-=2\pi^2 R^3 \f{\m(1+\m)^3}{(1-\n)^2} \sr{\m(1-\l^2)(\l-\n)},  \nn\\
&& \O_-^\psi=\f{1}{R}  \f{1-\n}{1+\m}   \sr{  \f{\l}{(1+\m)(1+\l)(\l-\n) }},  \nn\\
&& \Phi_-=\f{3\pi R}{2(2\pi)^{1/3}}\f{1+\m}{1-\n} \sr{\f{(1-\m)(\m+\n)(1-\l)}{\m}}.
\eea
With the quantities above, we can verify the relation
\be
S_+S_-=4\pi^2 J_\psi q^3,
\ee
in \cite{Castro:2012av}
and $T_+ S_+=T_-S_-$ proposed in \cite{Chen:2012mh}. The above relation suggests that there are  $J_\psi$ picture corresponding to the conserved angular momentum $J_\psi$ and possibly the dipole $q$ pictures to describe the dipole black ring holographically. Here we focus on the $J_\psi$  picture and discuss the existence of $q$ picture in Section~\ref{CONC}.

\subsection{CFT from thermodynamics}

It can be easily verified that there are Smarr formulas for the outer and inner horizons
based on Euler's law of homogeneous functions
\bea
&& M=\f{3}{2}(T_+ S_+ +\O_+^\psi J_\psi) +\f{1}{2}\Phi_+ q  \nn\\
&&\phantom{M}=\f{3}{2}(-T_- S_- +\O_-^\psi J_\psi) +\f{1}{2}\Phi_- q,
\eea
which are equivalent to the thermodynamics of the outer and inner horizons
\bea \label{e2}
&&d M=T_+ \dd S_+ +\O_+^\psi\dd J_\psi +\Phi_+ d q \nn\\
&&\phantom{\dd M}=-T_- \dd S_-+\O_-^\psi\dd J_\psi +\Phi_+ d q.
\eea
Following \cite{Chen:2012mh}, we define new quantities
\bea
&& \f{1}{T_{R,L}}=\f{1}{T_+} \pm \f{1}{T_-},  \nn\\
&& S_{R,L}=\f{1}{2}(S_+ \mp S_-),  \nn\\
&& \O_{R,L}^{\psi}=\f{\O_+^{\psi}/T_+  \pm  \O_-^{\psi}/T_-}{2/T_{R,L}},  \nn\\
&& \Phi_{R,L}=\f{\Phi_+/T_+  \pm  \Phi_-/T_-}{2/T_{R,L}},
\eea
we get the Smarr formulas and the first laws of the left- and right-moving sectors
\bea
&&M=3(T_R S_R +\O_R^\psi J_\psi) +\Phi_R  q \nn\\
&&\phantom{M}=3(T_L  S_L+\O_L^\psi J_\psi)+ \Phi_L  q,
\eea
\bea \label{e14}
&&\f{1}{2}d M=T_R  d S_R+\O_R^\psi d J_\psi +\Phi_R  dq \nn\\
&&\phantom{\f{1}{2} d M}=T_L  d S_L+\O_L^\psi d J_\psi+\Phi_L  dq.
\eea
Explicitly, we have
\bea
&& T_{R,L}=\f{\n(1+\n)}{4\pi R} \f{1}{(\m+\n)^{3/2}\sr{\l} \pm (1+\n)\m^{3/2}\sr{\l-\n} } \sr{\f{1-\l}{1+\l}} ,\nn\\
&& S_{R,L}=\pi^2 R^3 \f{(1+\m)^3\sr{1-\l^2}}{(1+\n)(1-\n)^2} \lt[ (\m+\n)^{3/2}\sr{\l} \mp (1+\n)\m^{3/2}\sr{\l-\n}  \rt],  \nn\\
&& \O_{R,L}^\psi=\f{1}{2 R} \f{1}{(1+\m)^{3/2}\sr{1+\l}}\f{ (\m+\n)^{3/2}\sr{\l-\n} \pm (1-\n^2)\m^{3/2}\sr{\l}  }
                                                        { (\m+\n)^{3/2}\sr{\l} \pm (1+\n)\m^{3/2}\sr{\l-\n} },   \\
&& \Phi_{R,L}=\f{3\pi R}{4(2\pi)^{1/3}} \f{1+\m}{1-\n} \f{(1-\n)(\m+\n)\sr{\m\l} \pm (1+\n)\m\sr{(\m+\n)(\l-\n)}}
                                             {(\m+\n)^{3/2}\sr{\l} \pm (1+\n)\m^{3/2}\sr{\l-\n}}
                                           \sr{(1-\m)(1-\l)}.  \nn
\eea

Setting $dq=0$ in (\ref{e2}), we get
\be
d J_\psi=T_L^\psi  d S_L  - T_R^\psi d S_R,
\ee
with the temperatures of dual CFT in the $J_\psi$ picture
\bea
T_{L,R}^\psi=\f{T_{L,R}}{\O_R^\psi-\O_L^\psi}=\f{(1+\m)^{3/2} \lt[ (\m+\n)^{3/2}\sr{\l} \pm (1+\n)\m^{3/2}\sr{\l-\n} \rt]}
                                                {4\pi \m^{3/2}(\m+\n)^{3/2}\sr{1-\l}}. \label{dipoleT}
\eea
From the Cardy formula
\be
S_{L,R}=\f{\pi^2}{3}c_{L,R}^\psi T_{L,R}^\psi,
\ee
we could derive the central charges
\be
c_L^\psi=c_R^\psi=\f{12\pi R^3(1-\l)\m^{3/2}(1+\m)^{3/2}(\m+\n)^{3/2}\sr{1+\l}}{(1+\n)(1-\n)^2}=6 q^3,
\ee
which is exactly the same central charges suggested in  \cite{Emparan:2004wy,Bena:2004tk,Cyrier:2004hj,Emparan:2008qn,Larsen:2005qr,Loran:2008mm,Goldstein:2011jh} for the extreme dipole black ring. Therefore, we conjecture that {\it a generic nonextremal dipole black ring (\ref{dipoler}) is holographically described by a 2D CFT with the central charges $c_L^\psi=c_R^\psi=6q^3$ and the temperatures (\ref{dipoleT})}.

In the extremal limit $\n=0$, the central charge and the temperature become
\bea \label{e3}
&& c_L^\psi=6 q^3,  \nn\\
&& T_L^\psi=\f{1}{2\pi} \lt( \f{1+\m}{\m} \rt)^{3/2} \sr{\f{\l}{1-\l}}.
\eea

\subsection{CFT from ASG analysis}

In this subsection, we reproduce the picture (\ref{e3}) from ASG analysis of the near horizon geometry of extremal dipole black ring. Firstly, we define new angles
\be
\td \phi=\f{2\pi}{\D \phi}\phi, ~~~ \td \psi=\f{2\pi}{\D \psi}\psi,
\ee
which have the period of $2\pi$. We also define $r=-1/y$, then the extremal ring has merging horizons at $r=0$. In the new coordinates $(t,r,x,\td\phi,\td\psi)$, the metric can be written as
\be
ds^2=-N^2 dt^2+g_{rr}dr^2+g_{xx}dx^2+g_{\td\phi\td\phi}d\td\phi^2+g_{\tps\tps} \lt( d\tps+N^\tps dt \rt)^2,
\ee
with
\bea
&& N^2=-g_{tt}+\f{g_{t\phi}^2}{g_{\phi\phi}}, ~~~ g_{rr}=\f{g_{yy}}{r^4},  \nn\\
&& g_{\tph\tph}=\lt( \f{\D\phi}{2\pi} \rt)^2 g_{\phi\phi}, ~~~ g_{\tps\tps}=\lt( \f{\D\psi}{2\pi} \rt)^2 g_{\psi\psi}, \nn\\
&& N^\tps=-\f{g_{t\psi}}{g_{\psi\psi}}\f{2\pi}{\D\psi}.
\eea
Following \cite{Chen:2011wm}, we expand the metric at the horizon
\bea
&& N^2=r^2 f_1^2+\ma O(r)^3,  \nn\\
&& g_{rr}=\f{f_2^2}{r^2} +\ma O\lt( \f{1}{r} \rt),  \nn\\
&& N^\tps=-\O_+^\tps +r f_3^\tps+\ma O(r)^2,  \nn\\
&& f_4=f_2/f_1, ~~~ f^\tps=f_3^\tps f_4.
\eea
Explicit calculation gives
\bea
&& f_1^2=\f{(1-\l)(1-\m x)(1+\l x)}{\m\l(1+\l)}, \nn\\
&& f_2^2=R^2\m^2(1-\m x)(1+\l x),   \nn\\
&& f_3^\tps=\f{1}{R}\f{1-\l}{\l}\f{1}{(1+\m)^{3/2}\sr{1+\l}},  \nn\\
&& f_4=R\m^{3/2}\sr{\f{\l(1+\l)}{1-\l}},  \nn\\
&& f^\tps=\lt( \f{\m}{1+\m} \rt)^{3/2}\sr{\f{1-\l}{\l}}.
\eea
We define the transformations
\be
t \to \f{f_4}{\e}t, ~~~ r \to \e r, ~~~ \tps \to \tps+\f{\O_+^\tps f_4}{\e}t,
\ee
and take the limit $\e \to 0$, and then we get the near horizon geometry of the ring
\be
ds^2=\a(x) \lt( -r^2 dt^2+\f{dr^2}{r^2} \rt)+ \b(x)dx^2  +\g_{\tph\tph} d\tph^2  +\g_{\tps\tps}(d\tps+f^\tps r dt)^2, \label{NHEDR}
\ee
with
\bea
&& \a(x)=f_2^2=R^2\m^2(1-\m x)(1+\l x),  \nn\\
&& \b(x)=g_{xx}|_{r=0}=\f{R^2\m^2(1-\m x)(1+\l x)}{1-x^2},  \nn\\
&& \g_{\tph\tph}=g_{\tph\tph}|_{r=0}=\f{R^2\m^2(1+\m)^3(1-\l)(1-x^2)}{(1-\m x)^2},  \nn\\
&& \g_{\tps\tps}=g_{\tps\tps}|_{r=0}=\f{R^2(1+\m)^3\l(1+\l)(1-\m x)}{\m(1+\l x)}.  \nn
\eea
As now there is no rotation along $\phi$, the near horizon geometry is simpler. For fixed $x$, it is a product of a warped AdS$_3$  and a $S^1$, with the warped AdS$_3$ being a AdS$_2$ with a U(1) fibre. The horizon area is just
\be
A_+=(2\pi)^2 \int^1_{-1}dx \sr{\b(x)\g_{\tph\tph}\g_{\tps\tps}}
   =8\pi^2 R^3\m^{3/2}(1+\m)^3\sr{\l(1-\l^2)}.
\ee
 For the geometry (\ref{NHEDR}), the U(1) symmetry in the fibre could be enhanced to a Virasoro algebra with a central charge. From \cite{Chen:2011wm}, we know the central charges and the temperature of the dual CFT
\bea
&& c_L^\tps= \f{3f^\tps A_+}{2\pi}=12\pi R^3 \m^3 (1+\m)^{3/2} (1-\l)\sr{1+\l}=6 q^3,  \nn\\
&& T_L^\tps=\f{1}{2\pi f^\tps}=\f{1}{2\pi} \lt( \f{1+\m}{\m} \rt)^{3/2} \sr{\f{\l}{1-\l}}.
\eea
This is in exactly agreement with the results (\ref{e3}) from the thermodynamics, and of course in match with the results in \cite{Emparan:2004wy,Bena:2004tk,Cyrier:2004hj,Emparan:2008qn,Larsen:2005qr,Loran:2008mm,Goldstein:2011jh}.

\section{Superradiant Scattering}\label{SC}

When we consider the scattering off an extreme Kerr black hole, the perturbations induce the infinitesimal excitations and the black hole is not exactly extreme and actually becomes near-extremal. Correspondingly the right-moving sector in the dual CFT gets excited, but its temperature is still infinitesimal small. In this situation, if the scattering is very near the superradiant bound, it could be understood as the scattering off the near-NHEK geometry. It is remarkable that the scattering amplitudes are in good agreements with the CFT predictions\cite{Bredberg:2009pv,Cvetic:2009jn,Hartman:2009nz,Chen:2010ni}. In this section, we consider the same issue in the context of black rings.
But actually our analysis is quite general and could be applied to all kinds of five-dimensional black rings and black holes that have regular outer and inner horizons.  We firstly show how to get the near horizon geometry of a near extremal black ring, and then analyze the scattering of a massless scalar under the geometry.

\subsection{Near horizon geometry of near extremal black ring}

We start from a general metric for a five-dimensional black hole or black ring
\bea\label{5D}
&&ds^2=-N^2 d\hatt^2+g_{rr}d\hatr^2+g_{xx}dx^2 +g_{\phi\phi}(d\hph+N^\hph d\hatt)^2 +g_{\psi\psi}(d\hps+N^\hps d\hatt)^2
\nn\\ && \phantom{ds^2=}
+2g_{\phi\psi}(d\hph+N^\hph d\hatt)(d\hps+N^\hps d\hatt).
\eea
Note that we have denote the coordinates as $(\hatt,\hatr,x,\hph,\hps)$. We consider the case when the black ring or black hole has two horizons located at $\hatr=r_\pm$, then we have
\bea
&&N^2=f_1^2(\hatr-r_+)(\hatr-r_-), \nn\\
&&g_{rr}=\f{f_2^2}{(\hatr-r_+)(\hatr-r_-)}, \nn\\
&&N^\hph=-\O_+^\hph+(\hatr-r_+)f_3^\hph +\ma O(\hatr-r_+)^2,  \nn\\
&&N^\hps=-\O_+^\hps+(\hatr-r_+)f_3^\hps +\ma O(\hatr-r_+)^2,
\eea
with $f_1,f_2$ being regular functions of $(\hatr,x)$ at the horizons $r_\pm$, and $\O_+^\hph,\O_+^\hps,f_3^\hph,f_3^\hps$ being constants. After imposing the nearly extremal condition
\be
r_+ - r_-=\e r_0,
\ee
with $r_0$ being finite and $\e$ being infinitesimally small, we get the Hawking temperature at the outer horizon
\be \label{e4}
\hat T_+=\f{\e r_0}{2\pi f_4},
\ee
with $f_4=\f{f_2}{f_1}|_{\hatr=r_+}$ being a finite constant. Following \cite{Bredberg:2009pv,Hartman:2009nz}, we may define the new coordinates as
\bea \label{e5}
&& \hatt=\f{f_4}{\e}t,  \nn\\
&& \hatr=r_++\e r,  \nn\\
&& \hph=\phi+ \lt( \f{\O_+^\hph f_4}{\e} +f^\phi r_0 \rt)t, \nn\\
&& \hps=\psi+ \lt( \f{\O_+^\hps f_4}{\e} +f^\psi r_0 \rt)t,
\eea
with $f^\phi=f_3^\hph f_4,f^\psi=f_3^\hps f_4$. We take the limit $\e \to 0$ and get the near horizon  geometry of a five-dimensional nearly extremal black ring
\bea \label{e7}
&& ds^2=\a(x) \lt[ -r(r+2r_0) dt^2+\f{dr^2}{r(r+2r_0)} \rt]+\b(x)dx^2  \nn\\
&& \phantom{ds^2=}
        +\g_{\phi\phi}(x)[d\phi+f^\phi (r+r_0) dt]^2+\g_{\psi\psi}(x)[d\psi+f^\psi (r+r_0) dt]^2  \nn\\
&& \phantom{ds^2=}
        +2\g_{\phi\psi}(x)[d\phi+f^\phi (r+r_0) dt][d\psi+f^\psi (r+r_0) dt].
\eea
The geometry can be viewed as a nonextremal black hole with the outer horizon located at $r=0$ with the Hawking temperature, and the angular momenta
\bea
&& T_+=\f{r_0}{2\pi},  \nn\\
&& \O_+^\phi=-f^\phi r_0,  \nn\\
&& \O_+^\psi=-f^\psi r_0.
\eea
Note that the Hawking temperature of the original geometry (\ref{e4}) $\hat T_+$ is infinitesimally small, but the temperature of the near horizon geometry $T_+$ is finite. This is due to the redefinition of the time (\ref{e5}), and in fact
\be
T_+=\f{f_4}{\e}\hat T_+.
\ee

When we consider the scattering off the geometry (\ref{5D}), the wave-function could be expanded in terms of $e^{-i\ho\hatt+i m_\hph \hph + m_\hps \hps }$ since the translation along $\hatt,\hph,\hps$ are isometries of spacetime and $\ho,m_\hph,m_\hps$ are well-defined quantum numbers. For the geometry (\ref{e7}), we may do similar expansion and
 make the identification
 \be
 e^{-i\ho\hatt+i m_\hph \hph + m_\hps \hps }=e^{-i\o t+im_\phi\phi+m_\psi\psi},
 \ee
 which gives us the relation between the original quantum numbers $(\ho,m_\hph,m_\hps)$ and the  new quantum numbers $(\o,m_\phi,m_\psi)$ in the near horizon geometry
\bea
&& m_\hph=m_\phi, ~~~ m_\hps=m_\psi,  \nn\\
&& \ho -m_\hph\O_+^\hph-m_\hps\O_+^\hps=\f{\e}{f_4} \lt( \o -m_\phi\O_+^\phi-m_\psi\O_+^\psi \rt).
\eea
If we choose $\o,m_\phi,m_\psi$ to be finite and in the superradiance region of near horizon geometry
\be
\o < m_\phi\O_+^\phi+m_\psi\O_+^\psi,
\ee
then we always have  $\ho,m_\hph,m_\hps$ near the superradiant bound of the original geometry
\be
\ho \simeq m_\hph\O_+^\hph+m_\hps\O_+^\hps.
\ee
It can be verified that the Boltzmann factors of the original and near horizon geometries are equal
\be
e^{-\f{\ho -m_\hph\O_+^\hph-m_\hps\O_+^\hps}{\hat T_+}}=e^{-\f{\o -m_\phi\O_+^\phi-m_\psi\O_+^\psi}{T_+}}.
\ee
If we identify it with that of CFT $e^{-\f{\o_L^\phi}{T_L^\phi}-\f{\o_L^\psi}{T_L^\psi}-\f{\o_R}{T_R}}$ where $\o_L^\phi=m_\phi,\o_L^\psi=m_\psi, \o_R=\o$, we get the CFT temperatures
\be \label{e6}
T_L^\phi=\f{1}{2\pi f^\phi}, ~~~ T_L^\psi=\f{1}{2\pi f^\psi}, ~~~ T_R=\f{r_0}{2\pi}.
\ee
This is in accord with the result in Kerr case\cite{Bredberg:2009pv,Hartman:2009nz}.

\subsection{Scattering amplitudes}

Let us consider the scattering of a massless scalar field in the near horizon geometry (\ref{e7}). The equation of motion for a massless scalar $\Phi$ is
\be
\f{1}{\sr{-g}}\p_\m \sr{-g}g^{\m\n}\p_\n \Phi=0.
\ee
Expanding the scalar as
\be
\Phi=R(r)X(x)e^{-i\o t+i m_\phi \phi+i m_\psi \psi},
\ee
we have the radial equation
\bea \label{e16}
&& \p_r r(r+4\pi T_R) \p_r R(r)
+\f {\lt( T_L^\phi T_L^\psi\o+T_L^\psi T_R m_\phi +T_L^\phi T_R m_\psi\rt)^2}{4\pi T_L^{\phi2}T_L^{\psi2}T_R r} R(r)  \nn\\
&&
-\f {\lt( T_L^\phi T_L^\psi\o-T_L^\psi T_R m_\phi -T_L^\phi T_R m_\psi\rt)^2}{4\pi T_L^{\phi2}T_L^{\psi2}T_R (r+2r_0)} R(r)
=\L R(r),
\eea
and the angular equation
\be
\f{\a}{\b} \p_x^2 X(x)+\f{2\a'\b\g-\a\b'\g+\a\b\g'}{2\b^2\g}\p_x X(x)
   -\f{\a}{\g} \lt( m_\psi^2 \g_{\phi\phi}+m_\phi^2\g_{\psi\psi}-2m_\phi m_\psi \g_{\phi\psi} \rt)X(x)=K X(x)
\ee
with the prime denoting $\p_x$, $\g=\g_{\phi\phi}\g_{\psi\psi}-\g_{\phi\psi}^2$, and the separation constants
\be
\L+K+\f{(m_\psi T_L^\phi+m_\phi T_L^\psi)^2}{4\pi^2 T_L^{\phi2} T_L^{\psi2}}=0.
\ee
Explicitly, for doubly rotating black ring, the angular equation is
\bea
&& \f{(1+r_+x)^2}{1-r_+^2}\p_x (1-x^2)\p_x X(x)
-\f{(1+r_+^2)(1+x^2)+4r_+ x}{2(1+r_+)^4(1+r_+ x)^2} \lt[ m_\psi^2 r_+^2 \phantom{\f{1}{2}} \rt. \nn\\
&& \lt.
-m_\psi m_\phi r_+(1+4r_++r_+^2)+m_\phi^2 \f{h(r_+,x)}{4(1-r_+)(1-x^2)} \rt]X(x)=K X(x),
\eea
with $h(r_+,x)$ defined as (\ref{h}), and for dipole black ring it becomes
\be
\p_x(1-x^2)\p_x X(x)- \f{1+\l x}{(1+\m)^3} \lt[ m_\psi^2 \f{\m^3(1+\l x)}{\l(1+\l)}  +m_\phi^2 \f{(1-\m x)^3}{(1-\l)(1-x^2)} \rt] X(x)
=K X(x).
\ee
Even though the scalar equations in ordinary black ring background is impossible to be variable-separated \cite{Emparan:2006mm},  the scalar equations are separable in the near horizon geometry.

Solving the radial equation (\ref{e16}), we get the solution with the ingoing boundary condition at the horizon
\be
R(z) \sim z^{-i\g}(1-z)^h F(a,b;c;z),
\ee
with the definition
\bea
&& z=\f{r}{r+4\pi T_R}, \nn\\
&& h=\f{1}{2} \pm \sr{\f{1}{4}+\L},  \nn\\
&& \g=\f{m_\phi}{4\pi T_L^\phi}+\f{m_\psi}{4\pi T_L^\psi}+\f{\o}{4\pi T_R}, \nn\\
&& a=h-i\f{\o}{2\pi T_R},  \nn\\
&& b=h-i\f{m_\phi}{2\pi T_L^\phi}-i\f{m_\psi}{2\pi T_L^\psi}, \nn\\
&& c=1-2i\g.
\eea
As $r\to\inf$, we have
\be
R(r) \sim A r^{h-1}+B r^{-h}
\ee
with
\be
A=\f{\G(2h-1)\G(c)}{\G(a)\G(b)},
~~~B=\f{\G(1-2h)\G(c)}{\G(c-a)\G(c-b)}.
\ee
After identifying the CFT conformal weights and charges as
\be \label{e22}
h_L=h_R=h, ~~~ \o_L^\phi=m_\phi, ~~~ \o_L^\psi=m_\psi,~~~\o_R=\o.
\ee
We get the retarded Green function and absorption cross section as \cite{Chen:2010ni}
\bea \label{e23}
&& G_R \sim \f{B}{A} \sim \sin[\pi(h_L+i\f{\o_L^\phi}{2\pi T_L^\phi}+i\f{\o_L^\psi}{2\pi T_L^\psi})]
              \sin[\pi(h_R+i\f{\o_R}{2\pi T_R})] \nn\\
&& \phantom{G_R \sim}\times
              \lt| \G (  h_L+i\f{\o_L^\phi}{2\pi T_L^\phi}+i\f{\o_L^\psi}{2\pi T_L^\psi} )  \rt|^2
              \lt| \G (  h_R+i\f{\o_R}{2\pi T_R )}  \rt|^2,                                             \\
&& \s \sim \textrm{Im}G_R \sim  \sinh \lt(  \f{\o_L^\phi}{2 T_L^\phi}+\f{\o_L^\psi}{2 T_L^\psi}+\f{\o_R}{2 T_R} \rt) \nn\\
&&\phantom{\s \sim \textrm{Im}G_R \sim}\times
              \lt| \G (  h_L+i\f{\o_L^\phi}{2\pi T_L^\phi}+i\f{\o_L^\psi}{2\pi T_L^\psi} )  \rt|^2
              \lt| \G (  h_R+i\f{\o_R}{2\pi T_R )}  \rt|^2.
\eea
Remembering that
\be
 \f{\o_L^\phi}{T_L^\phi}+\f{\o_L^\psi}{T_L^\psi}+\f{\o_R}{T_R}=\f{\o -m_\phi\O_+^\phi-m_\psi\O_+^\psi}{T_+},
\ee
we see the effect of superradiance.

At first looking, the above relation is not consistent with the CFT prediction, as there
seems to be two left-temperatures in the expression. However, it should be understood in an appropriate way. Firstly, for the two black ring cases we have been discussing, they are in good agreement with CFT predictions, as there is only one possible CFT dual. Actually in doubly rotating case there is only $J_\phi$-picture with $1/T_L^\psi=0$ such that the above relation could be simplified. Similarly for the dipole black ring, there is only $J_\psi$ picture with $1/T_L^\phi=0$. Secondly, for the possible general black ring, 5D Kerr black hole, or uplifted Kerr-Newmann black hole, the interpretation of the above relations should be careful. Recall that in discussing the scattering amplitudes of low-frequency scattering, the hidden conformal symmetry only become manifest when we turn off some quantum numbers and keep only one of them, the angular momentum or the $U(1)$ charge. This suggests that when we let $m_\psi$ or $m_\phi$ vanishing,  we have scattering amplitudes in well match with the CFT predictions in $J_\phi$ or $J_\psi$ picture. On the other hand, the existence of both quantum numbers are permitted. In this case, the dual picture is related to the $J_\phi$ or $J_\psi$ picture by an $SL(2,Z)$ transformation\cite{Chen:2011wm,Chen:2011kt}.
More explicitly, the $SL(2,Z)$ symmetry of the CFT duals comes from the redefinition of the angles
\be
\lt( \ba{c} \phi \\ \psi \ea \rt)
\to
\lt( \ba{c} \phi' \\ \psi' \ea \rt)=
\lt(\ba{cc} a& b \\c & d \ea \rt)
\lt( \ba{c} \phi \\ \psi \ea \rt),
\ee
with
\be \lt(\ba{cc} a& b \\c & d \ea \rt) \in SL(2,Z). \ee
The invariance of the combination $m_\phi \phi+m_\psi \psi$ requires that
\be
\lt( \ba{cc} m_\phi & m_\psi \ea \rt)
\to
\lt( \ba{cc} m_{\phi'} & m_{\psi'} \ea \rt)=
\lt( \ba{cc} m_\phi & m_\psi \ea \rt)
\lt(\ba{cc} d & -b \\ -c & a \ea \rt).
\ee
In the $SL(2,Z)$ symmetry, the temperatures of the two CFT dual pictures transform as
\be
\lt( \ba{c} 1/T_L^\phi \\ 1/T_L^\psi \ea \rt)
\to
\lt( \ba{cc} 1/T_L^{\phi'} \\ 1/T_L^{\psi'} \ea \rt)=
\lt(\ba{cc} a& b \\c & d \ea \rt)
\lt( \ba{c} 1/T_L^\phi \\ 1/T_L^\psi \ea \rt).
\ee
Remembering (\ref{e22}) $\o_L^\phi=m_\phi,\o_L^\psi=m_\psi$, we see that the combination $\f{\o_L^\phi}{T_L^\phi}+\f{\o_L^\psi}{T_L^\psi}$, which appears in the retarded Green function and absorption cross section (\ref{e23}), is invariant under the $SL(2,Z)$ transformation, viz. that
\be
\f{\o_L^\phi}{T_L^\phi}+\f{\o_L^\psi}{T_L^\psi}=\f{\o_L^{\phi'}}{T_L^{\phi'}}+\f{\o_L^{\psi'}}{T_L^{\psi'}}.
\ee
Setting $\o_L^{\phi'}$ or $\o_L^{\psi'}$ vanishing, we get the the retarded Green function and absorption cross section of CFT dual in the $\psi'$ or $\phi'$ picture . Note that for the near extremal black hole or black ring, the right-moving frequency $\o_R$ and temperature $T_R$ are the same for both $\phi$ and $\psi$ pictures, and is invariant under the $SL(2,Z)$ transformation.

\section{Conclusion and Discussion\label{CONC}}

In this paper we discussed the holographic CFT duals for two kinds of black rings, i.e.\! the neutral doubly rotating black ring and the dipole charged black ring. We first calculated the dual CFT central charge and temperatures of the extremal doubly rotating black ring, and found that it agrees with the extremal limit of the ones got in \cite{Chen:2012mh} using the thermodynamic method. Moreover, we investigated the CFT dual for the dipole black ring from different points of view and found consistent picture.
 Furthermore, we discussed the superradiant scattering off the extreme black rings and found the scattering amplitudes to be in good match with CFT predictions. As a byproduct, we got the superradiant scattering amplitudes for the 5D Kerr and uplifted 5D Kerr-Newman black holes, which shows the $SL(2,Z)$ symmetry.

The study in this paper shows that the thermodynamics method is an effective way in setting up the nonextremal black hole/CFT correspondence. Up to now, we only know that it works well by scrutinizing concrete examples, but we do not know why it works from the first principle. Definitely it needs further intense study. In particular, it would be nice to check our conjecture on non-extremal black rings by some other ways.

It seems that for every charge that appears at the first law of thermodynamics there is a corresponding CFT dual. For example, if we have the first laws at the outer and inner horizons
\bea
&& dM=T_+ d S_++\Phi_+ dQ +\cdots \nn\\
&& \phantom{dM}=-T_- d S_-+\Phi_- dQ +\cdots ,
\eea
with $Q$ being a general charge and $\Phi_\pm$ being corresponding potentials. The first laws could be written as
\bea
&& \f{1}{2}dM=T_L d S_L+\Phi_L dQ +\cdots \nn\\
&& \phantom{\f{1}{2}dM}=T_R d S_R+\Phi_R dQ +\cdots ,
\eea
As long as $\Phi_L\neq\Phi_R$, we could set the other charges be zero and get the temperatures of CFT in the dual $Q$ picture. If this is the case, then from (\ref{e14}), there seems be a dipole picture for the CFT dual for the dipole black ring. But we have no idea how to set up the dipole picture from conventional method. This is a question that deserves further research. Also, there are other kinds of charged black rings, and it would be nice to see how the CFT duals behave using the thermodynamic method and how these duals be derived by conventional methods.

\vspace*{10mm}
\noindent {\large{\bf Acknowledgments}}

The work was in part supported by NSFC Grant No.~10975005.

\vspace*{5mm}



\end{document}